\documentclass[sigconf,screen,authorversion]{acmart}
\usepackage{url}
\usepackage{multirow}
\usepackage{tabularx}
\usepackage{enumitem}

\usepackage{pgfplots}
\pgfplotsset{width=8.5cm,compat=1.9}

\RequirePackage[skins]{tcolorbox}
\tcbuselibrary{skins,xparse,breakable}
\newtcolorbox{custombox}[1]{
    breakable,
	colback=gray!10,
	colframe=gray!50,
	left=1mm,
	right=1mm,
	top=1mm,
	bottom=1mm,
	fonttitle=\bfseries,
	arc=0mm,
	leftrule=1mm,
	rightrule=0mm,
	toprule=0mm,
	bottomrule=0mm,
	notitle,
	before=\par\smallskip\noindent,
	before upper={\textbf{#1 } },
}

\AtBeginDocument{%
  \providecommand\BibTeX{{%
    \normalfont B\kern-0.5em{\scshape i\kern-0.25em b}\kern-0.8em\TeX}}}

\copyrightyear{2024} 
\acmYear{2024} 
\setcopyright{acmlicensed}\acmConference[ITiCSE 2024]{Proceedings of the 2024 Innovation and Technology in Computer Science Education V. 1}{July 8--10, 2024}{Milan, Italy}
\acmBooktitle{Proceedings of the 2024 Innovation and Technology in Computer Science Education V. 1 (ITiCSE 2024), July 8--10, 2024, Milan, Italy}
\acmDOI{10.1145/3649217.3653580}
\acmISBN{979-8-4007-0600-4/24/07}

\begin{document}

\title[Enhancing Student Engagement in Large-Scale Capstone Courses]{Enhancing Student Engagement in Large-Scale Capstone Courses: An Experience Report}

\author{Asma Shakil}
\affiliation{%
 \institution{The University of Auckland}
 \city{Auckland}
 \country{New Zealand}}
 \email{asma.shakil@auckland.ac.nz}

\author{Paul Denny}
\affiliation{%
 \institution{The University of Auckland}
 \city{Auckland}
 \country{New Zealand}}
 \email{paul@cs.auckland.ac.nz}

\renewcommand{\shortauthors}{Asma Shakil \& Paul Denny}

\begin{abstract}
Computer science (CS) capstone courses offer students a valuable opportunity to gain hands-on experience in software development, practice essential soft skills, and enhance their employability prospects. They are a core component in many CS undergraduate degrees and address the ACM curricula requirements of inculcating professional dispositions in students and making them aware of the broader societal implications of computing.  However, coordinating a capstone course, especially for a large student cohort, can be a daunting task for academic staff.  It demands considerable time and energy for planning and coordinating activities between students, academic staff, and any external stakeholders. In this experience report, we outline the iterative development and refinement of our capstone course as it grew substantially in size over a span of six consecutive sessions. We outline the pedagogies that helped us to enhance student engagement and motivation in the course as assessed by end-of-course surveys and students' written reflections. We share the lessons that we have learnt and provide recommendations to educators who are designing new capstone courses or looking to scale existing ones.
\end{abstract}

\ccsdesc[500]{Social and professional topics~Computing education}

\keywords{computer science education, experiential learning, capstone course}

\maketitle

\section{Introduction}
The importance of developing ``professional dispositions'' in CS graduates, defined as ``cultivable behaviors desired in the workplace'', have been greatly emphasised in the competency models of the latest CS 2023 curriculum put forward by the joint task force of ACM, IEEE-CS, and AAAI \cite{CS2023_2023}. Capstone courses serve as an ideal point to help students practice these professional dispositions before they transition into higher studies or workplace. Capstone courses aim to help students integrate the knowledge gained in their degree program, reflect on prior learning and hone their professional skills \cite{carter2011ideas, keller2011employ}. With these aims in mind, capstone courses in computer science generally require students to work in teams on substantial software development projects which are often based on real-world problems. Students get the opportunity to practice their technical skills while also developing several essential soft skills such as communication, teamwork, time-management, and self-directed learning which help improve their employability prospects \cite{Khakurel2020effect, li2023software}.

However, coordinating capstone courses is a formidable task for academic staff, especially for large cohorts \cite{clear2001resources}. It requires considerable time, effort, and energy to coordinate course activities such as seeking project proposals, liaising with external stakeholders, organizing showcase events, evaluating group work, and engaging with students at a deeper level than is needed in more traditional lecture-based courses.  As capstone teachers in the School of Computer Science at The University of Auckland for six consecutive iterations, we experienced many of these challenges when we scaled our course from 36 students in Semester 1 2021, to 220 in the recent most offering in Semester 2, 2023. 

In this experience report, we present the design and iterative refinement of our capstone course as it has scaled to include a significantly increased student cohort. We adopted several pedagogies that focused on developing professional skills in students such as teamwork, communication, project management, peer assessment, self-reflection, and a willingness to learn new technologies. We took several measures to enhance student motivation and engagement such as providing them the opportunity to work on real-life projects pitched by clients external to the course, collaborating with industry partners for infrastructure support and training, incorporating recognition such as awards for high performance, and extending the outreach of the capstone project outcomes to a broad audience including prospective employers, clients, and future students. These measures helped us to not only maintain but enhance student engagement as the course grew significantly in enrolments over the last six iterations.
 
We share our journey of successfully scaling our capstone course, reflecting on the positive experiences and identifying areas for improvement. We also provide a set of concrete recommendations for CS capstone instructors. We hope that by presenting our reflections of this experience, and sharing the lessons that we have learned along the way, we can help educators who are designing new capstone courses or looking to scale existing ones. 

\section{Related Work}

Computer science capstone courses provide students with a culminating experience in which their theoretical study is complemented  with practical skills that prepare them for professional careers and help students to connect their learning with modern software development concepts and practices \cite{sherriff2018capstones, paasivaara2017do}. \citeauthor{tenhunen2023systematic}  \cite{tenhunen2023systematic} provide a recent systematic literature review of 127 articles on capstone courses in software engineering and computer science published between 2007–2022. They propose a taxonomy of course features based on ACM/IEEE guidelines for capstone courses. 

A common capstone course model includes teaching software engineering theories in lectures while allowing students to apply these theories in large group projects \cite{dugan2011survey}. Capstone courses emphasize learning essential professional skills including communication, teamwork, and ethics \cite{oleary2017innovative}. \citeauthor{herbert2018reflections} \cite{herbert2018reflections} emphasized the importance of team-based capstone projects delivered by external clients with the goal of developing professional skills and increasing employability. Carter \cite{carter2011ideas} proposed focusing on soft skills like communication and teamwork within the capstone curriculum. These authors agree on the value of team-based projects, emphasizing the significant role of capstones in developing students' soft skills and enhancing their readiness for the professional world. 

The use of capstone projects in real-world applications has also been highlighted. Bloomfield et al. \cite{bloomfield2014service} shared a successful implementation of a Service Learning Practicum (SLP) course sequence intended to be a year-long capstone for computer science seniors. In this work the students developed software for local non-profit organizations resulting in good outcomes for the participating parties.  Similarly, Murphy et al. \cite{murphy2017two} presented a two-course sequence involving ``real projects for real customers,'' recognizing the benefits of students improving code they didn't write themselves.  Focusing more on pedagogical strategies for teaching capstones, Heckman et al. \cite{heckman2018ten} shared insights from a software engineering course at North Carolina State University, highlighting the importance of working with large existing systems and teaching assistant support for course success. They also highlight the importance of constantly updating course content to stay in line with modern practices. 

In summary, capstone courses provide an opportunity for students to combine theoretical knowledge with practical real-world skills to develop important professional attributes. However, capstone course coordination is challenging due to factors such as managing external clients, seeking project sources, team formation and organization, preparation of course resources, and group work evaluation. Interestingly, very little research has explored the intrinsic motivation factors for teaching capstone courses and their influence on teaching approaches \cite{hixson2012capstone}. Moreover, strategies to foster student engagement and investment in projects is essential, as these are directly linked to project outcomes \cite{parker2017how}. There exists a significant opportunity for educators to share their experiences and build a community of capstone coordinators in computer science.

\section{Context}

At The University of Auckland, a new capstone course requirement was introduced by the Faculty of Science in 2019 to fix the curriculum fragmentation issue identified in our Bachelor of Science degree program. The course was aimed at helping students assess and reflect on their whole-of-program learning so that they can `join the dots' in their knowledge. The course was also seen as an opportunity to help students develop transferable skills between and within disciplines, all of which are essential attributes of our graduate profile\footnote{https://www.auckland.ac.nz/assets/science/study-with-us/docs/graduate-profile/bsc-graduate-profile-2019.pdf}, and in great demand by employers \cite{osmani2019graduates}.

The CS capstone course was launched as a pilot in our School in 2020 as an elective course for final-year students. The pilot run of the course was supervised by one academic staff and had 18 students in Semester 1 and 17 in Semester 2. In semester 1, 2021, the course became a standard elective course and the first author was made the teacher and coordinator of the course. In semester 2, 2021 the course changed to a core (i.e. required) course and the cohort size grew from 36 to 121 (Table \ref{tab:course_numbers}). Since then, the course has grown steadily in size with the latest offering having 220 students. Our teaching team comprises teachers (academic staff on teaching track), tutors (postgraduate students) and a director (academic staff on research track). The teachers are responsible for teaching and coordinating the course and evaluating students' work, tutors oversee students' progress and provide feedback on a weekly basis, and the course director helps in strategic planning. The growth in teaching team size can be seen in Table \ref{tab:course_numbers}.

\begin{center}
\begin{table}
\caption{Capstone course offerings and growth in numbers}
  \label{tab:course_numbers}
\begin{tabular}{||l l l c c c c ||} 
 \hline
 Year & Sem & Students & Teams & Teachers & Tutors & Clients\\ 
 \hline\hline
 \multirow{2}{*}{2020} & S1 & 18 & 4 & 1 & 0 & 0 \\ 
 \cline{2- 7} 
  & S2 & 17 & 3 & 1 & 0 & 0\\
 \hline
  \multirow{2}{*}{2021} & S1 & 36 & 7 & 1 & 1 & 5 \\ 
 \cline{2- 7} 
  & S2 & 121 & 17 & 1 & 3 & 13\\
 \hline
  \multirow{2}{*}{2022} & S1 & 152 & 27 & 1 & 3& 11 \\ 
 \cline{2- 7} 
  & S2 & 202 & 37 & 2 & 3& 17\\
 \hline
  \multirow{2}{*}{2023} & S1 & 112 & 20 & 1 & 3& 10\\ 
 \cline{2- 7} 
  & S2 & 220 & 39 & 2 & 4& 17\\
 \hline
\end{tabular}
\end{table}
\end{center}

\section{STUDENT FEEDBACK} \label{secn:student_feedback}
Our course design and evolution over the span of six iterations has been informed by student feedback 
from the following sources:
\begin{itemize}[leftmargin=*]
\item{\textit{Formal end-of-course evaluations} : Our university conducts comprehensive course evaluations at the end of each semester, using Likert scale and open-ended questions. Open-ended responses from these surveys have been valuable in providing us with feedback on course strengths and areas for improvement.}
\item{\textit{Students' reflection documents} : We ask students to write reflection documents at the end of the semester (Table \ref{tab:course_assessment}). These documents are a rich source for us to analyze students' experience, learning, team collaboration, and overall course satisfaction.}
\end{itemize}

\noindent This feedback underpins many of our course decisions as detailed in Sections \ref{secn:course_structure} and \ref{secn:reflections}.

\section{Capstone Course Evolution} \label{secn:course_structure}
We use the framework proposed in the systematic literature review on capstone courses by \citeauthor{tenhunen2023systematic} \cite{tenhunen2023systematic} to describe the development and iterative refinement of our course structure. The framework is based on five elements: course duration (one semester or a full year), team size (individual to large groups), client and project origin (course staff or external), project implementation (software prototypes, reports, technologies used), and assessment methods (final grading and continuous guidance and assessment).

\subsection{Duration}
\begin{figure*}
  \centering 
  \includegraphics[width=0.9\textwidth]{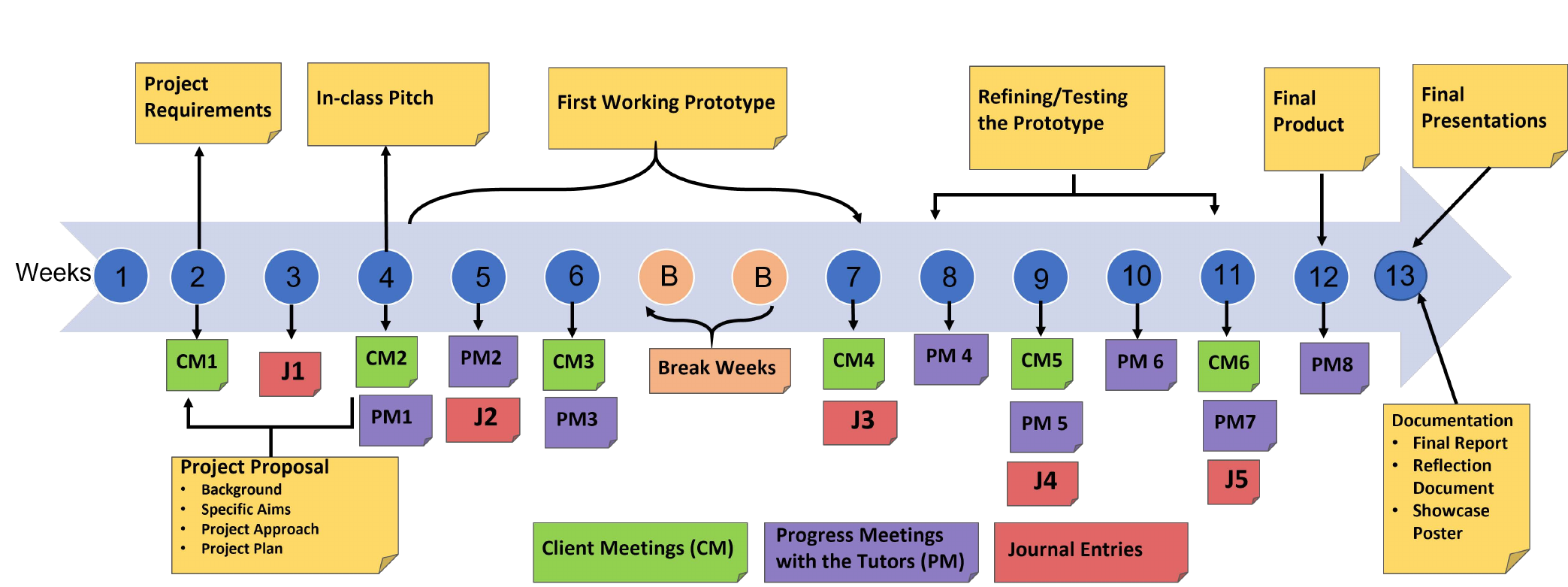}
    \caption{A timeline of our course showing the artifacts produced and client and progress meeting instances.} 
  \label{fig:course_timeline} 
\end{figure*}

Our capstone course is a semester-long course that runs for 14 weeks with a two-week mid-semester break in between (Figure \ref{fig:course_timeline}). Students are required to put in 10 hours worth of work per week. It is offered twice a year in both Semester 1 and Semester 2. Given the large size of our graduating cohort (500+ students), it is difficult for us to have a single year-long capstone course as recommended by ACM/IEEE \cite{CS2013}. A one-semester capstone duration is the practice adopted by most institutions \cite{tenhunen2023systematic}.

\subsection{Team size and Formation} 
When we expanded our course from 36 students in S1, 2021 to 121 in S2, 2021, we increased team sizes from 5--6 to 7--8 members to manage workload. However, this proved ineffective as 50\% of students reported in our end-of-semester formal course evaluation that, \textit{``the large team size made it difficult to delegate tasks and work effectively together''}. Consequently, we have reverted to 5-6 member teams, regardless of class size.

Our pre-course survey data shows that students are concerned about being in ineffective teams (see data in Figure \ref{fig:fears}). Studies show no difference in student experiences between student-formed and instructor-formed teams \cite{hastie2019design}. In our first course iteration (S1, 2021) with 35 students, we formed diverse teams based on grades, skills, and experience, but found this approach to be unscalable. In the next semester (S2, 2021), with 121 students, we initially let students form their own teams, then assigned remaining students to teams based on project interest. This led to some students being placed in an already formed team. The open-ended responses in our formal end-of-semester evaluations revealed that students in pre-formed teams \textit{felt unwelcome}. We now allow self-team formation followed by random grouping of remaining students, ensuring equal footing in new teams.

\begin{figure}
  \centering 
\begin{tikzpicture}
\begin{axis}[
    ybar=2pt,
    bar width=5pt,
    enlarge x limits=0.15,
    legend style={at={(0.5,-0.3),
    cells={align=left}
    },
      anchor=north,
      legend columns=-1, 
        },
    ylabel={Percentage of Responses},
    symbolic x coords={Difficult Project,Ineffective Team,Boring Project,Presentations,Time Management},
    xtick=data,
    x tick label style={rotate=20,anchor=east, font=\footnotesize},
    title={\textit{What do you fear the most in the course?}},
    title style={yshift=-0.2cm, xshift=0cm}, 
    ytick={0,10,20,30,40,50},
    height=5cm, 
    ]
\addplot coordinates {(Difficult Project,26) (Ineffective Team,44.5) (Boring Project,6) (Presentations,13) (Time Management,8)};
\addplot coordinates {(Difficult Project,24.5) (Ineffective Team,45) (Boring Project,6.7) (Presentations,15.8) (Time Management,8.7)};
\addplot coordinates {(Difficult Project,26.3) (Ineffective Team,47.4) (Boring Project,6.5) (Presentations,13) (Time Management,7)};
\addplot coordinates {(Difficult Project,22.3) (Ineffective Team,47.6) (Boring Project,7) (Presentations,10) (Time Management,9.7)};
\legend{{S1, 2022} , {S2, 2022},{S1, 2023},{S2, 2023}, \\ {N= 66}, {N= 66},{N= 66},{N= 66}}
\end{axis}
\end{tikzpicture}
 \caption{Students' response to the pre-course survey} 
  \label{fig:fears} 
\end{figure}

\begin{figure}
  \centering 
\begin{tikzpicture}
\begin{axis}[
    ybar=2pt,
    bar width=5pt,
    enlarge x limits=0.1,
    legend style={at={(0.5,-0.5)},
      anchor=north,
      legend columns=-1, 
        },
    ylabel={Percentage of Responses},
    symbolic x coords={Interesting Project,Technical Skills,Soft Skills,Teamwork Experience,Independent \& Peer Learning, Challenging Myself!},
    xtick=data,
    x tick label style={rotate=20,anchor=east, font=\footnotesize},
    title={\textit{What are you looking forward to the most in the course?}\\\textit{(Select all that apply.)}},
    title style={yshift=-0.2cm, xshift=0cm, align=center}, 
    ytick={0,10,20,30,40,50,60,70,80,90},
    height=5cm, 
    ]
\addplot coordinates {(Interesting Project,85) (Technical Skills,81.2) (Soft Skills,63) (Teamwork Experience,72) (Independent \& Peer Learning,46) (Challenging Myself!,39.5)};
\addplot coordinates {(Interesting Project,88.6) (Technical Skills,80) (Soft Skills,62.9) (Teamwork Experience,69.3) (Independent \& Peer Learning,42.9) (Challenging Myself!,38.6)};
\addplot coordinates {(Interesting Project,96.4) (Technical Skills,82.1) (Soft Skills,60.8) (Teamwork Experience,64.3) (Independent \& Peer Learning,50) (Challenging Myself!,37.5)};
\addplot coordinates {(Interesting Project,87.8) (Technical Skills,75.6) (Soft Skills,65) (Teamwork Experience,70) (Independent \& Peer Learning,45.5) (Challenging Myself!,41.5)};

\legend{{S1, 2022}, {S2, 2022},{S1, 2023},{S2, 2023},{N= 66}, {N= 66},{N= 66},{N= 66}}
\end{axis}
\end{tikzpicture}
 \caption{Students' response to the pre-course survey} 
  \label{fig:lookforwardto} 
\end{figure}

\subsection{Clients and Project Ideas}
The initial pilot runs of the course in 2020 did not involve external stake holders. Instead, students worked on self-pitched projects. Since 2021, we have emphasized real-world experience in the course by involving external clients. Our clients include industry partners, local businesses, and staff from the different units and faculties within our University. Our project ideas are now derived from clients' needs and are thus based on real-world problems. This shift from student-initiated projects to client-driven, real-world software solutions aligns with ACM/IEEE guidelines \cite{CS2013, CC2020} and enriches student learning \cite{sherriff2018capstones,dunlap2005problem}, but it is a complex and time-intensive endeavor not widely adopted by many institutions \cite{schorr2020experience, bastarrica2017can, judith2003client}.
We solicit projects for our course by sending emails about four weeks prior to the start of the semester. Each semester we get between 20 -- 25 projects. We then vet through all the projects to ensure that they have a software development component that is doable within the 14-week duration of our course. Students bid for their top five projects as a team and we allow up to four teams to work on projects that receive multiple bids.

\subsection{Project Implementation}
Given our bi-weekly client meeting structure (Figure \ref{fig:course_timeline}), our capstone teams follow an agile development methodology with sprint cycles of two weeks. Students showcase their progress to clients every two weeks and refine their prototypes based on feedback. 

\subsubsection{Artifacts:}
There are seven artifacts that students produce in our course (Table \ref{tab:course_assessment} and Figure \ref{fig:course_timeline}). The major artifact is the project build or software prototype. Other artifacts include an individual reflection document, a group report, a final presentation, and a showcase poster. Depending on client needs and project specifics, some teams may also need to create additional deliverables like user manuals or code documentation.

\subsubsection{Project Phases:}
Our capstone students go through several phases of the  software development life cycle including requirements elicitation, planning, designing , developing and testing. The students meet with clients biweekly (Figure \ref{fig:course_timeline}) which allows them to gradually understand client needs and refine the project accordingly. Each session is an opportunity to present progress, receive feedback, and adjust the project's direction, ensuring it evolves in alignment with the client's vision. Some teams work on projects that are a continuation from previous semesters. In such cases, teams get to experience aspects of code maintenance and refactoring as well.

\begin{center}
\begin{table}
\caption{Course Artifacts and Assessment Weights}
  \label{tab:course_assessment}
\begin{tabular}{||l l l c||} 
 \hline
 Artifact & Weight & Type & Week Due \\ 
 \hline\hline
 \multirow{2}{*}{Project Proposal} & 7\% (Document) & \multirow{2}{*}{Group} & \multirow{2}{*}{4} \\ 
 \cline{2- 2} 
  & 3\% (Oral Pitch) & & \\
 \hline
  \multirow{2}{*}{Progress Updates} & 5\% (Journals) & \multirow{2}{*}{Individual} & \multirow{2}{*}{4} \\ 
 \cline{2- 2} 
  & 5\% (Meetings) & & \\
 \hline
  {Project Build} & 40\% & Group & 12\\ 
  \hline
   {Reflection Document} & 15\% & Individual& 13\\ 
  \hline
   {Group Report} & 10\% & Group& 13\\ 
    \hline
   {Presentation} & 10\% & Individual& 13\\ 
    \hline
   {Showcase Poster} & 5\% & Individual& 13\\ 
 \hline
\end{tabular}
\end{table}
\end{center}

\subsubsection{Project Technologies:}

Our capstone projects encompass a variety of CS topics including development of web and mobile applications, digital games, Python packages, AI models, simulations, and desktop applications (Figure \ref{fig:capitalise}). We advise students to discuss technology stacks with their clients in the initial meeting. Our experience shows that most clients permit teams to choose their own technologies. 

We use GitHub for collaborative coding, setting up a class organization where students have administrative rights for independent project management, aligning with recommendations for GitHub use in advanced courses \cite{tu2022github}. Additionally, students must employ a project management tool like Jira, Notion, or Trello for project planning and task management. We require access to their project management sites and a link in their GitHub README file.

\subsection{Course Assessments}
We use both end-of-semester and continuous progress monitoring to assess our students. There are seven assessed components in our course (refer to Table \ref{tab:course_assessment}).
\subsubsection{End of course student assessment}
The end of course assessments comprise of the project source code (worth 40\% of the marks), group report (worth 10\%), reflection document (worth 15\%), final presentation (worth 10\%), and showcase poster (worth 5\%). The project proposal, which includes a written document (worth 7\%) and an oral in-class pitch (worth 3\%) is assessed in week 4 of the semester (Figure \ref{fig:course_timeline} \& Table \ref{tab:course_assessment}).

\subsubsection{Continuous student assessment and guidance}
We assess continuous progress through three mechanisms -- weekly progress meetings with the teams, fortnightly journals submitted individually by each student, and at the biweekly client meetings (Figure \ref{fig:course_timeline}). The weekly progress meetings are worth 5\% of the marks and are conducted by the tutors in the course. The meetings are modeled on the stand-up meeting format in SCRUM-based capstone courses \cite{chang2022adapting, jimenez2016scrum}, where each team member needs to answer three questions, \textit{``What did you do in the last week?''}, \textit{``What will you do in the coming week?''}, and \textit{``Are there any impediments to your progress?''}. Students also write weekly journals detailing their activities, analyzing their experiences, reflecting on their learning, outlining plans for the upcoming week, and assessing their team's performance (Table \ref{tab:journal_template}). These journals, each contributing 1\% to the grade, are submitted fortnightly (Figure \ref{fig:course_timeline}). As capstone instructors, we attend all biweekly client-team meetings to continuously monitor team progress and offer support as needed.

\begin{center}
\begin{table}
\caption{An outline of the template that we provide to students for their weekly journal entries.}
  \label{tab:journal_template}
\begin{tabularx}{\columnwidth}{|X|X|}
 \hline
 Category & Guiding Question \\ 
 \hline\hline
 A description of the activities  & What did I do this week? \\ 
 \hline
  Analysis of the experience  & What were the most significant events this week? How did I feel during this week? \\ 
   \hline
  Articulation of Learning  & What did I learn this week? \\ 
  \hline
  Planning  & What will I do next week? \\
  \hline
  Team Dynamics’ Assessment & How are the behavioural relationships between members? \\
 \hline
\end{tabularx}
\end{table}
\end{center}

\section{Reflections and Recommendations} \label{secn:reflections}
We reflect on our experience of conducting six consecutive sessions of our course (post-pilot) from S1, 2021 to S2 2023  during which we have coordinated 147 teams (involving 843 students) and liaised with 57 different clients across 72 projects (see Table \ref{tab:course_numbers}). We present our reflections as a set of lessons we have learned and recommendations for the computing education community.
    
\subsection{Student Engagement and Motivation}
Our pre-course survey results have shown that most students (about 60 -- 65\%) enjoy teamwork and are excited about the course. However, there is always a significant percentage that have either not worked in teams before (\(\sim \)10--12\%) or do not enjoy teamwork (\(\sim \)25--30\%) due to poor team experience in prior courses. Such students are anxious about the capstone course.

To help ease students' concerns and boost their motivation and engagement with the course, we used the Expectancy-Value Theory (EVT) framework \cite{eccles2002motivational}. The EVT is a psychological framework in which \textit{`expectancy'} refers to an individual's belief about how likely they are to succeed in a task while \textit{`values'} relate to how much an individual values the task at hand. Values include -- \textit{attainment value} (the personal importance of doing well on the task), \textit{intrinsic value} (the interest in the task itself), \textit{utility value} (the practical benefits of completing the task), and \textit{cost value} (the negative aspects of engaging in the task, such as time and effort required). 

We use a pre-course survey to gauge and positively influence students' \textit{expectancy} for success, by asking them about their anticipations and concerns going into the course (see Figures \ref{fig:fears} and \ref{fig:lookforwardto}). This approach aligns with the `expectancy' aspect of the EVT model, as it directly addresses students' expectations for success. We see these sentiments in the responses to our end-of-course evaluations:

\begin{center}
\textit{``Prior to starting this project, I had never done such a large scale coding project. Consequently, I was excited to undertake this particular project, as it would give me the opportunity to gain significant practical experience in this area.''}
\end{center}

\begin{center}
\textit{``The experience of doing the capstone project gave me a greater appreciation of teamwork. Prior to this course, I had limited teamwork experience. This project stood out as this team worked very closely together over the semester, compared to just a few weeks like some of my other projects.''}
\end{center}

We emphasize the `attainment value' by  reminding students of the course's role in enhancing employability prospects and making a real-world impact. In our first lecture, we show several job advertisements that are currently open for new graduates and explain how the course will help students to demonstrate the skills that employers are looking for in these advertisements. The `intrinsic value' is fostered by allowing students to choose projects that interest them, and providing creative freedom in tackling open-ended, real-world problems. 
`Utility value' is achieved by involving external clients who pitch real-world projects, thereby increasing the practical application and relevance of their work. Lastly, we address the `cost value' by emphasizing the importance of effective time and project management. We feel that this holistic approach, grounded in EVT theory has helped to considerably enhance student engagement and motivation in our course as assessed by end-of-course evaluations (Table \ref{tab:motivation_scores}) and students' final reflection documents, e.g.,:

\begin{center}
    \textit{``Going into the capstone, I had pessimistic thoughts about whether I would even pass it, let alone enjoy it or if I would have a good team. As a completed capstone student, I look back to this feeling and wish I could have reassured myself that it would be the best course, client, and team experience I have ever had.''}
\end{center}
\begin{center}
    \textit{``I enjoyed this course a lot, and I think it has  motivated me to attempt to do more projects in my spare time. I have gained a better understanding of how to work with and for a client in a software development environment.''} 
\end{center}
\begin{center}
    \textit{``Having reached the end of a long, stressful, but overall very enjoyable semester, I can safely say that this course has been a highlight of my degree. I've had a great time getting to know my team members, and an even better time working on a project that has real value''}. 
\end{center}
\begin{center}
    \textit{``Our team chose to undertake this project because it actually had real-world impact and that if we were successful, it would actually go to good use as opposed to existing in a repo and doing nothing!''}.
\end{center}

\begin{center}
\begin{table}[ht]
\caption{End-of-course survey results for the last six iterations showing the percentage of students who \textit {Generally Agreed} (\%GA) (i.e., either \textit{Agreed} or \textit{Strongly Agreed})}
  \label{tab:motivation_scores}
\begin{tabular}{|p{0.55\columnwidth} | p{0.1\columnwidth} p{0.1\columnwidth} p{0.1\columnwidth}|}
 \hline
 Question & Year & Sem & \%GA\\ 
 \hline\hline
\multirow{6}{*}{\small{ {\begin{tabular}[c]{@{}l@{}} \textit{I could stay motivated and engaged} \\ \textit{with my learning in the course.} \\ \end{tabular}} }}  & \multirow{2}{*}{2021} & S1 & 71.4\%  \\ 
 \cline{3- 4} 
  & & S2 & 70\% \\
 \cline{2- 4} 
  & \multirow{2}{*}{2022} & S1 & 90\% \\ 
 \cline{3- 4} 
  & & S2 & 76.7\%  \\
 \cline{2- 4} 
  & \multirow{2}{*}{2023} & S1 & 87.5\% \\ 
 \cline{3- 4} 
  & & S2 & 94.4\% \\
 \hline
\end{tabular}
\end{table}
\end{center}

\begin{custombox}
{Recommendation: Use pre-course surveys to gauge and shape students' expectations of success.} These surveys help to clarify course objectives, set expectations and alleviate concerns. Also, using real job ads as examples, demonstrate how the course will help develop skills that employers are looking for in fresh graduates.
\end{custombox}

\begin{custombox} {Recommendation: Incorporate a diverse range of real-world projects in the capstone course.} Using authentic, real-world problems significantly boosts student motivation and performance. Allowing students to select projects that resonate with their interests fosters a deeper sense of ownership and dedication, thereby enhancing their engagement and outcomes.
\end{custombox}

\subsection{Awards and Recognition}
\begin{figure}
  \centering 
  \setlength{\fboxsep}{0pt}%
  \fbox{\includegraphics[width=\columnwidth]
  {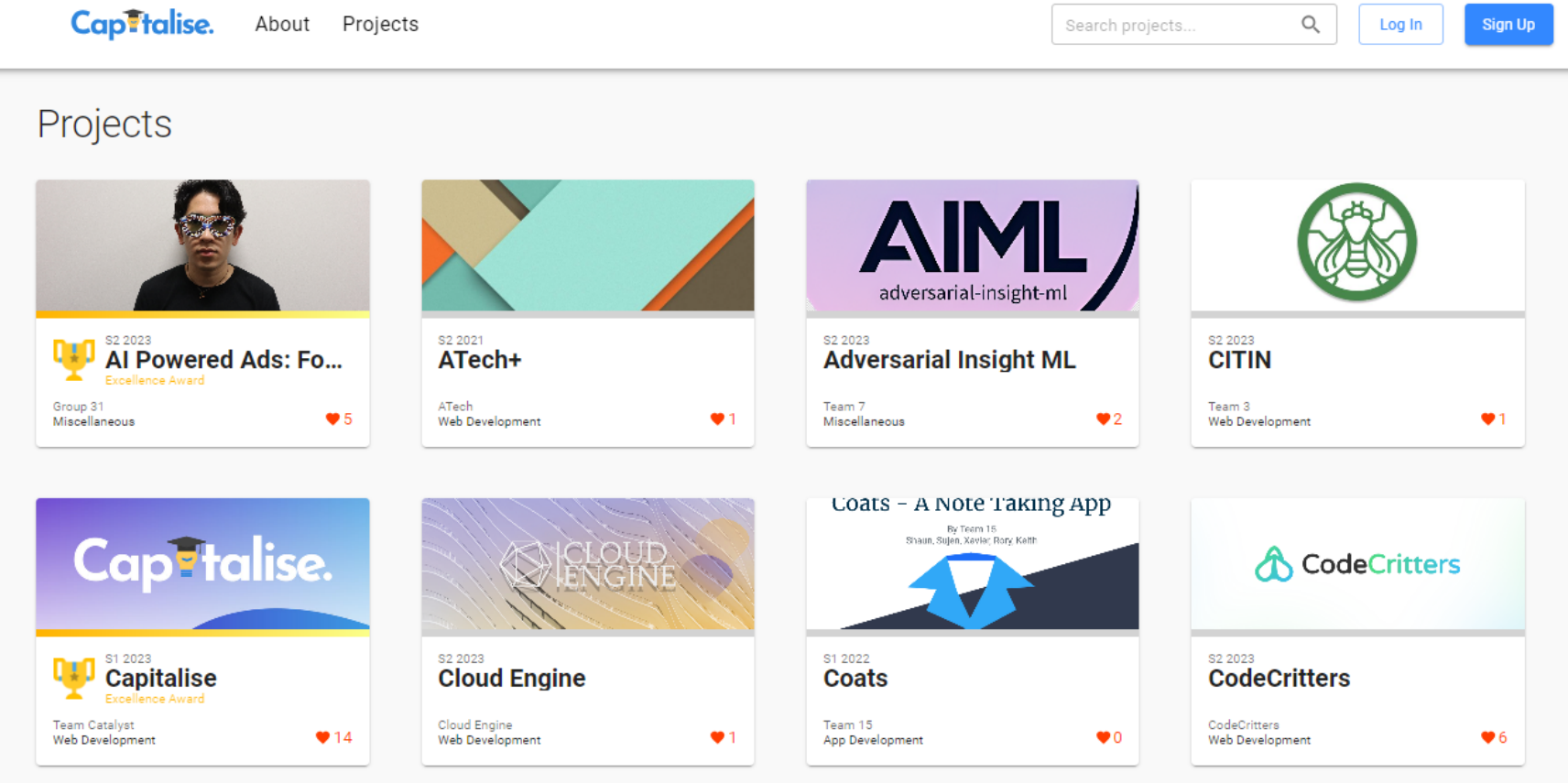}}
    \caption{An online portfolio to showcase capstone projects.} 
  \label{fig:capitalise} 
\end{figure}

We hold a public event at the end of each semester where students showcase their capstone project outcomes to a wide audience which includes industry representatives, academics from the university, their friends and family members. We give out several awards to our high-performing teams at the event which include an \textit{Excellence Award}, a \textit{Community Impact Award} and a \textit{People’s Choice Award}. The nominations for the first two awards are based on client feedback and overall performance in the course. These  awards are sponsored in part by our School and in part by our industry partners. We have also developed an online website to showcase students' work to prospective clients and employers (Figure \ref{fig:capitalise}). Interestingly, this website was developed as a project in our capstone course. 
\begin{custombox} {Recommendation: Incorporate awards and extend the outreach of capstone project outcomes.} Motivate students to excel by offering opportunities to present their projects to potential employers, gain recognition at a public ceremony, and network with industry professionals. 
\end{custombox}

\subsection{Groupwork Evaluation} \label{secn:group_evaluation}
Groupwork evaluation is a challenging task in any course \cite{fellenz2006toward}. We use several measures to ensure fair grading and penalize free-riders. Fair evaluation is crucial for student satisfaction given the high stakes the course has on group work (Table \ref{tab:course_assessment}). We use TeamMates\footnote{https://teammatesv4.appspot.com/web/front/home} to collect peer assessment data in four rounds --- post-proposal in week 5, and weeks 8, 10, and 13 (Figure \ref{fig:course_timeline}). The first three rounds are formative feedback and do not influence marks. However, they provide valuable insights to students about how their contributions are being weighed by their teammates and puts on alert any free-riders in the team. The final peer evaluation is a summative evaluation where students distribute a fixed number of points among their teammates based on their contribution to each group work component (Table \ref{tab:course_assessment}). We compute an average peer score (excluding self scores and outliers for fairness) and use it to adjust the team's mark for each student individually. We also ask students to recommend a grade for each teammate in their final reflection documents and an explanation justifying their recommendation. Since the reflection documents are submitted at the end of the semester and are confidential, students are very honest in the evaluations of their team mates and are eager to point out any free-riders who were not contributing as much as expected. We have found these written reflections to be very useful in triangulating peer scores. Finally, we also look at GitHub commit history and progress meeting records in cases where students complain that they have been wrongly evaluated by their teammates. 

\begin{custombox} {Recommendation: Ensure fair and accurate grading for groupwork through a multi-faceted evaluation approach.}  Use peer evaluations to adjust team grade for individual students based on their contribution. To guarantee fairness, these peer assessments should be corroborated with additional evidence, such as written reflections, analysis of each student's GitHub commit history, and continuous progress assessments.
\end{custombox}
\subsection{Capstone Coordination Effort} 
We dedicate more time and effort to the capstone course compared to other courses that we teach. The capstone effort includes soliciting clients and projects before the start of the semester, attending biweekly client meetings, assessing individual contribution to group work and organizing the end-of-semester showcase event. Our drive comes from seeing our students' enthusiasm and satisfaction with the course. Our School won the \textit{2022 Teaching \& Learning award} for \textit{providing a fun and engaging experience to the students in the capstone course.}  This acknowledgment of our hard work, much like the award system for students in our course, motivates us to continue to improve our course and provide an enriching culmination experience for our students. We also have a Capstone Champions group that comprises capstone instructors across all schools and departments in our faculty. We provide support to each other, share resources, and meet often to discuss strategies, tools, and techniques that we can apply in our own capstone courses. 

\begin{custombox}{Recommendation: Build a community of capstone instructors for support.} A community of instructors who teach capstone courses can go a long way in providing moral support and ease the burdens of capstone course coordination.  
\end{custombox}

\section {Conclusion and Future Work}
Coordinating a large capstone course can be an overwhelming task for academic staff. In this paper, we have shared our experience of successfully running a large capstone course over six consecutive iterations. The course which started as an elective course in our School in 2019 has evolved into a fully fledged core course that is taken by all our graduating students. Despite the large cohort size, we have been able to provide a fun, enriching, and valuable experience to our students as is evident in their feedback and end-of-semester reflections. Our hope is that by sharing our reflections and recommendations in this experience report, we can assist others who are looking to design new capstone courses or scaling existing ones.
Going forward, we plan to use the course as a platform to strengthen our School's ties with the local technology industry. We are excited to see how these partnerships will provide our graduating students the opportunities, the skill sets and the confidence they need as they transition into industry careers. 

\balance

\bibliographystyle{ACM-Reference-Format}
\bibliography{sample-base}

\end{document}